\documentclass[aps,pre,twocolumn,showpacs,floatfix]{revtex4}

\usepackage{graphicx, subfigure}
\usepackage{graphics}
\usepackage{latexsym}
\usepackage{subfigure}
\usepackage{multirow}
\usepackage{amsmath}
\usepackage{amssymb}
\usepackage{txfonts}
\usepackage[usenames]{color}

\newcommand{\be}{\begin{equation}}
\newcommand{\ee}{\end{equation}}
\newcommand{\bea}{\begin{eqnarray}}
\newcommand{\eea}{\end{eqnarray}}

\begin{document}

\title{Effects of intrinsic fluctuations in a prototypical chemical oscillator:  metastability and switching}
\author{C. Michael Giver, Bulbul Chakraborty}
\affiliation{Martin A. Fisher School of Physics, Brandeis University, Waltham, MA USA}

\date{\today}
\pacs{05.40.-a, 02.50.Ey, 82.40.Bj}

\begin{abstract}
Intrinsic or demographic noise has been shown to play an important role in the dynamics of a variety of systems including predator-prey populations, intracellular biochemical reactions, and oscillatory chemical reaction systems, and is known to give rise to oscillations and pattern formation well outside the parameter range predicted by standard mean-field analysis. Initially motivated by an experimental model of cells and tissues where the cells are represented by chemical reagents isolated in emulsion droplets, we study the stochastic Brusselator, a simple activator-inhibitor chemical reaction model. Our work builds on the results of recent studies and looks to understand the role played by intrinsic fluctuations when the timescale of the inhibitor species is fast compared to that of the activator. In this limit, we observe a noise induced switching between small and large amplitude oscillations that persists for large system sizes ($N$), and deep into the non oscillatory part of the mean-field phase diagram.  We obtain a scaling relation for the first passage times between the two oscillating  states . From our scaling function, we show that the first passage times have a well defined form in the large $N$ limit. Thus in the limit of small noise and large timescale separation a careful treatment of the noise will lead to a set of non-trivial deterministic equations different from those obtained from the standard mean-field limit.

\end{abstract}
 
\maketitle

\section{Introduction \label{sec-intro}}
In recent years systems with demographic or intrinsic fluctuations have gained an increasing amount of attention. The fluctuations in these systems arise from the stochastic nature of discrete reactions between a finite number of elements. Careful treatment of the intrinsic noise has lead to the discovery of many interesting behaviors not found in the traditional mean-field treatment of such problems. McKane and Newman showed that demographic noise can induce noisy oscillations, or quasicycles, where mean-field predicts a stable stationary state in a simple predator-prey model \cite{McKane:2005hl}. Butler and Goldenfeld then extended this to show that a similar mechanism can lead to noisy Turing patterns, quasipatterns, well outside the region of parameter space predicted by standard Turing analysis \cite{Butler:2011gh}.  Butler's work gives a possible solution to the fine tuning problem for pattern formation and provides a possible explanation for the ubiquity of Turing-like patterns observed in nature. The effects of intrinsic stochasticity has also been looked at in models of chemical oscillators and cell signaling and intracellular biochemical reactions, all of which show interesting non mean-field behavior \cite{Boland:2008,Biancalani:2010,Biancalani:2011jk,Artyomov27112007}.

In this paper we are interested in systems in which the dynamics of one variable is fast with respect to the other variables in the system.   This timescale separation can lead to excitability, meaning that if the system is sufficiently perturbed from a stable fixed point, it will go on a large excursion in phase space before returning to the fixed point. Examples of excitable systems include neuronal systems, and chemical oscillators such as the Belouzov-Zhabotinsky reaction \cite{Cross-Greenside}. In recent years, it has been recognized that, in the presence of  a large time scale separation, even small noise can lead to dynamical patterns that are absent in the mean-field model \cite{LeeDeVille:2006hg,Muratov:2008di,Muratov:2007vf}.   It is known that in systems with well-separated time scales, a singular Hopf bifurcation can occur leading to a large jump in oscillation amplitude and a transition to relaxation oscillations \cite{Baer:1986}.  The appearance of stochastic resonance has been demonstrated in a Brusselator close to a Hopf bifurcation \cite{Osipov:2000tr}. This stochastic resonance occurs when the dynamics of the activator species is fast compared to the inhibitor.   In this paper, we study the effects of intrinsic fluctuations in the opposite limit where the inhibitor species has the fast dynamics.  Our results show that , in this regime, the Brusselator switches between small and large amplitude oscillations with switching times that remain finite in the limit of large systems (small intrinsic noise).  The switchability is a property of the intrinsically noisy system and is not present in the meanfield model.

The remainder of this paper is structured as follows. In the next section we will present our version of the Brusselator model starting with the discrete reactions. Then we will briefly describe the behavior of the Brusselator in the mean-field limit. This section will conclude with a short discussion of the van Kampen system size expansion of the Brusselator model, which was included for completeness. For a more detailed discussion one should see references \cite{McKane:2004ir,Boland:2008,vanKampen:1992}. In section \ref{sec-Results} we will present the results of our simulations and analysis, all of which will be summarized in section \ref{sec-Discussion}.
\section{The Brusselator model \label{sec-Brusselator}}

The Brusselator, put forth by Ilya Prigogine in the 1960s, is a prototypical model of an autocatalytic chemical reaction system showing limit cycle behavior \cite{Prigogine:1971}. This model has been studied extensively over the past several decades including works on the mean-field limit, subject to external fluctuations, and more recently with noise due to intrinsic or demographic noise \cite{Boland:2008,Osipov:2000tr,Pena:2001ch,Vanag:2003dv,Biancalani:2011jk}. The Brusselator model consists of two species, an activator $X$ and inhibitor $Y$, which undergo the set of four reactions
\begin{eqnarray}
0 & \xrightarrow{N} &X \nonumber \\
X &\xrightarrow{1} &0\nonumber \\
X &\xrightarrow{b} &Y \nonumber\\
2X+Y & \xrightarrow{c}  &3X.
\label{eqn-reactions}
\end{eqnarray}
We assume the system is coupled to some external bath where $X$ particles are fed into the system with rate $N$ and leave with rate $1$, which set the size and timescale for our system respectively, and thus the system is maintained out of equilibrium. The parameter $b$ is the rate of exchange from $X$ to $Y$, while $c$ is the rate of the autocatalytic reaction converting $Y$ back to $X$. As will be shown below, $c$ also determines the separation of timescales between the dynamics of the two species $X$ and $Y$.\\
\begin{figure}
	\includegraphics[width=\columnwidth]{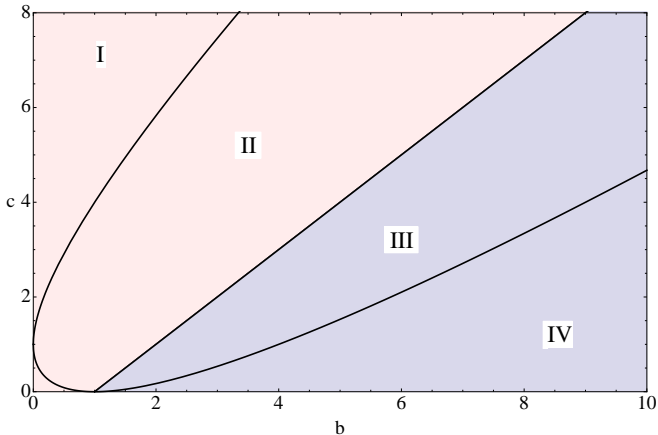}
	\caption{Phase diagram of the mean-field Brusselator. The Brusselator has a stable Fixed point in regions $I$ and $II$ and a stable limit cycle about an unstable fixed point in regions $III$ and $IV$. The eigenvalues of the Jacobian matrix are complex in regions $II$ and $III$, making the fixed point a stable, or unstable, spiral, while the eigenvalues in regions $I$ and $IV$ are real. }
	 \label{LinearPhaseDiagram}
\end{figure}
\subsection{Mean-Field - Linear Stability Analysis}

Using the law of mass action, we can write down a set of mean-field rate equations for the number density of the two species
\begin{eqnarray}
\label{eqn-MFx}
\dot{x} &= &1 - (1+b)x + c x^2 y\\
\dot{y} &= &bx - c x^2 y
\label{eqn-MFy}
\end{eqnarray}
where $x = X/N$ and $y = Y/N$. These equations have a unique fixed point at $x^* = 1, y^* = b/c$ that is linearly stable for $b < c+1$ and becomes unstable, giving rise to a stable limit cycle as the control parameter, $b$, is varied past the critical value $b_c = c +1$.

The phase diagram, to linear order, is shown in Fig. \ref{LinearPhaseDiagram}. In regions $I$ and $II$ the fixed point is stable, and in $III$ and $IV$ the fixed point is unstable and surrounded by a stable limit cycle. The eigenvalues of the Jacobian matrix in regions $II$ and $III$ are complex making the fixed point an unstable (or stable) spiral.
\subsection{Mean-Field - Beyond Linear Stability}

\begin{figure*}
	\includegraphics[width=\textwidth]{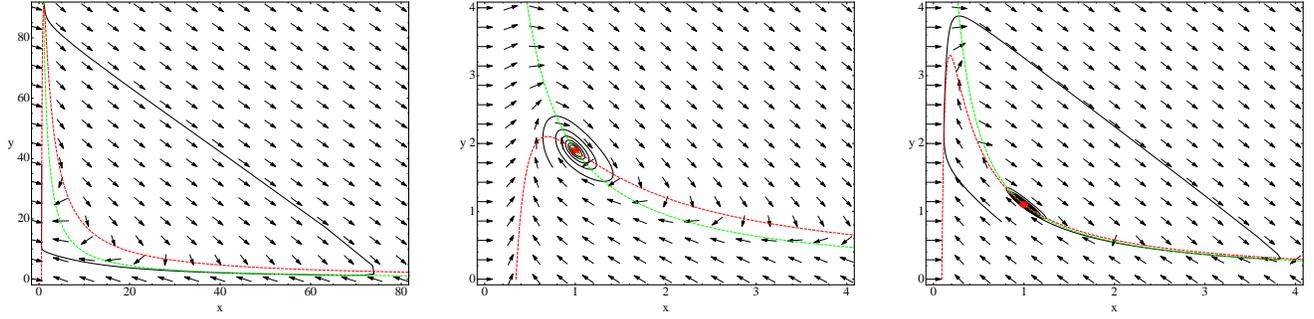}
		\caption{Phase space vector plot of Brusselator equations with nullclines(dashed lines) and example trajectory(solid line) starting near the fixed point in three different timescale regiemes; $c << 1$(left), $c = 1.0$(center), and $c > >1.0$(right).}
	\label{fig-MFPaseSpacePlots}
\end{figure*}

In the presence of a large time scale separation many oscillatory systems will become excitable, meaning that small perturbations from the stable fixed point or small amplitude limit cycle can send the system on a large amplitude excursion through the phase space. This time scale separation can be written explicitly by the change of variables $x^\prime = x$ and $y^\prime = c y$ in Eqs. \ref{eqn-MFx} and \ref{eqn-MFy}. Carrying out this substitution gives
\begin{eqnarray}
\label{eqn-MFx2}
\dot{x^\prime } &= &1 - (1+b)x^\prime +  x^{\prime2} y^\prime\\
\tau_{y} \dot{y^\prime} &= &bx^\prime  - x ^{\prime2} y^\prime
\label{eqn-MFy2}
\end{eqnarray}
where $\tau_{y} = c^{-1}$ is the time scale of the variable $\bar{y}$. In the large $c$ limit, the inhibitor has the fast dynamics whereas as $c \rightarrow 0$, the activator becomes the fast variable.  Within mean field theory, there is a marked change in behavior as the ratio of timescales is varied, as illustrated in Fig. \ref{fig-MFPaseSpacePlots}.   The vector plots of the phase space along with the nullclines, at $c = 0.01, c  = 1.0$ and $c = 9.0$ and $b = 0.9, b = 1.9$, and $b = 9.9$ below the bifurcation threshold, show that as $c \rightarrow 0$, the fixed point approaches the peak of the $X$-variable nullcline, and becomes pinned at that point.  As $c \rightarrow \infty$, on the other hand, the segments of the $X$ and $Y$ nullclines that lie to the right of the peak, approach each other and merge converting the single fixed point to a line of fixed points.  Sample trajectories shown in Fig. \ref{fig-MFPaseSpacePlots} demonstrate that when the time scales are comparable, the trajectory will simply spiral into the fixed point.  When $c$ is large, the trajectory will escape to the nullcline of the slow variable ($X$), which it will follow until it quickly jumps to the other slow branch  following that and ultimately spiraling into the fixed point along the incipient small limit cycle. The second slow branch is almost superposed on the $Y$ nullcline, and becomes the line of fixed points in the $c \rightarrow \infty$ limit.  When $c$ is small, as in Fig. \ref{fig-MFPaseSpacePlots}a,  the trajectory escapes along a $X=-Y$ path, then quickly jumps to the slow branch following it to the stable fixed point.

\begin{figure}
	\includegraphics[width=\columnwidth]{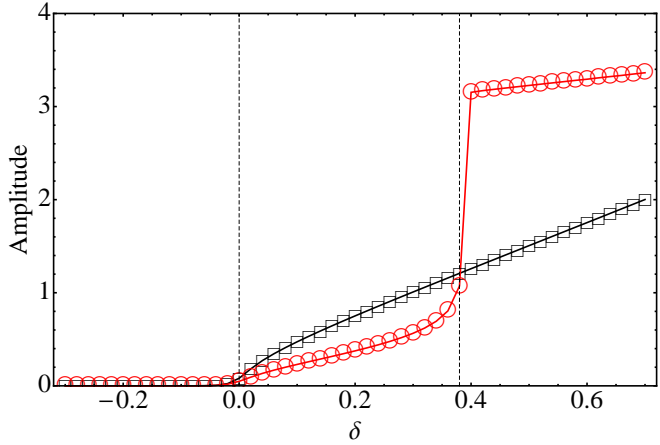}
	\caption{Amplitude of oscillation as b is varied past the bifurcation point when $x$ and $y$ have the same time scale, $c = 1.0$ ($\square$), and when $y$ is the fast variable, $c = 9.0$ ($\bigcirc$). Amplitudes were taken from numerical solutions to Eqs.  \ref{eqn-MFx} and \ref{eqn-MFy}. }
	 \label{fig-Amplitude}
\end{figure}

In all three cases, as the control parameter $b$ is tuned past the bifurcation point $b_c = c+1$ the fixed point becomes unstable and gives rise to a stable limit cycle with an amplitude that grows as $\sqrt{b - b_c}$ near $b_c$. In the regime where $c \gg 1$ the amplitude of the limit cycle will jump sharply as it is increased to the point where the trajectories can escape. At this point, the large excursion becomes the stable limit cycle. Fig. \ref{fig-Amplitude} shows the amplitude of oscillation as a function of the distance from the bifurcations, $\delta = b - b_c$.  It should be emphasized that in the mean field description the Brusselator is alway a monostable system being either at a fixed point, a small amplitude limit cycle or a large amplitude limit cycle.  As we show below, the presence of noise induces metastability and switching.

\subsection{Master Equation Formulation}

The rate equations analyzed in the previous section are of a mean field nature \cite{Boland:2008}, which ignores the demographic noise present in the intrinsically stochastic process described by Eq. \ref{eqn-reactions}.  In order to account for the stochasticity,  the ``random walk'' in chemical space is written down in the form of a Master equation \cite{McKane:2004ir,Boland:2008,vanKampen:1992}.   The relevant equation for the Brusselator is:
\begin{eqnarray}
\dot{P}(n_x,n_y,t) &= &[(\epsilon_x^{-}-1)T_1(n_x,n_y) \nonumber \\
& & +(\epsilon_x^{+}-1)T_2(n_x,n_y) \nonumber \\
& & +(\epsilon_x^{+}\epsilon_y^{-}-1)T_3(n_x,n_y) \nonumber \\
& & +(\epsilon_x^{-}\epsilon_y^{+}-1)T_4(n_x,n_y)  ]P(n_x,n_y,t).
\label{eqn-MasterEquation}
\end{eqnarray}
Here $P(n_x,n_y,t)$ is the time dependent probability of finding the system in the state $(n_x,n_y)$ at time $t$ and $\epsilon^{\pm}$ is a step operator that acts on functions of $n_x(n_y)$ by raising or lowing $n_x(n_y)$ by one, as follows:
\begin{eqnarray}
\epsilon_x^{\pm}f(n_x,n_y) & = & f(n_x\pm 1,n_y)\\
\epsilon_y^{\pm}f(n_x,n_y) & = & f(n_x,n_y\pm 1).
\end{eqnarray}
and the transition rates, $T_i(n_x,n_y)$ are given by
\begin{eqnarray}
T_1(n_x,n_y ) &= &N \nonumber \\
T_2(n_x,n_y ) &= &n_x \nonumber \\
T_3(n_x,n_y ) &= &bn_x \nonumber \\
T_4(n_x,n_y ) &= &\frac{c n_x(n_x-1)n_y}{N^2}.
\label{eqn-TransitionRates}
\end{eqnarray}

While the full Master equation cannot be exactly solved, significant insight can be gained by carrying out a system size expansion. We first assume that the number density of each species is given by the average density plus fluctuations of order $N^{-1/2}$, 
\begin{eqnarray}
\frac{n_x}{N} = x(t) + N^{-1/2}\xi_x(t)\\
\frac{n_y}{N} = y(t) + N^{-1/2}\xi_y(t).
\end{eqnarray}
It can be shown, that $x(t)$ and $y(t)$ satisfy the mean field equations presented in the previous section \cite{Boland:2008}.  Plugging the expansion into the Master equation and substituting the probability distribution of the fluctuations, $\Pi(\xi_x,\xi_y,t)$, for $P(n_x,n_y,t)$, we can expand the right hand side of Eq. \ref{eqn-MasterEquation} in powers of $N^{-1/2}$ and collect terms of the same order. From the leading order terms in the expansion we regain the mean-field equations given by Eqs. \ref{eqn-MFx} and \ref{eqn-MFy}. The next highest order terms give a linear Fokker-Planck equation for distribution of fluctuations, which is known to have a Gaussian solution. The Fokker-Planck equation can be equivalently described by the linear Langevin equation
\begin{equation}
\dot{\mathbf{\xi}} = K \mathbf{\xi}(t) + \mathbf{f}(t).
\label{eqn-Langevin}
\end{equation} 
Here we have written the Langevin equation in vector form where $\mathbf{\xi} = (\xi_x,\xi_y)$, $K$ is the drift matrix and is exactly equal to the Jacobian matrix  from the linear stability analysis of Eqs. \ref{eqn-MFx} and \ref{eqn-MFy}, and $\mathbf{f}(t) = (f_x,f_y)$ is Gaussian white noise described by
\begin{equation}
\langle f_i(t) f_j(t')\rangle = 2 D_{ij}\delta(t -t')
\end{equation}
The matrices $K$ and $D$ depend upon the mean field solutions, $x(t)$ and $y(t)$.   Evaluated at the fixed point,  these are  by:
\begin{equation}
K = 
 \begin{pmatrix} b-1 & c \\ -b & -c \end{pmatrix} 
\end{equation}

\begin{equation}
D = 
 \begin{pmatrix} 1+b & -b \\ -b & b \end{pmatrix} .
\end{equation}
From here one can easily obtain an expression for the power spectra of fluctuations of $X$ and $Y$, as was shown in \cite{Boland:2008}. The power spectra are given by
\begin{equation}
P_x(\omega) = \frac{2((1+b)\omega^2+c^2)}{(c-\omega^2)^2+(1+c-b)^2\omega^2}
\label{eqn-PowerSpecX}
\end{equation}
\begin{equation}
P_y(\omega) = \frac{2b((\omega^2+1+b)}{(c-\omega^2)^2+(1+c-b)^2\omega^2}.
\label{eqn-PowerSpecY}
\end{equation}

The power spectra calculated from the van Kampen approximation are peaked at non-zero frequencies for parameters in the fixed point regime of the mean-field Brusselator model. Here the system acts analogously to a damped driven pendulum. The intrinsic fluctuations drive the system at all frequencies, thus exciting the systems natural frequency. These oscillations are the quasicycles originally reported in \cite{McKane:2005hl}. 
\section{Numerical Results \label{sec-Results}}
\begin{figure}
	\includegraphics[width=\columnwidth]{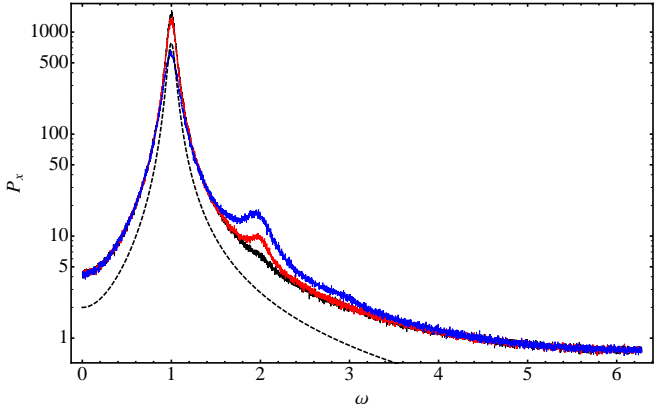}
	\caption{Power spectrum of the stochastic Brusselator averaged over $500$ realizations with $b = 1.9$ and $c = 1.0$ for system sizes $N = 10^3$(blue), $N = 10^4$(red), and $N = 10^5$(black) against the power spectrum calculated from the van Kampen expansion of the Master equation(dashed).}
	 \label{fig-PowerSpectrumC1}
\end{figure}
\begin{figure}
	\includegraphics[width=\columnwidth]{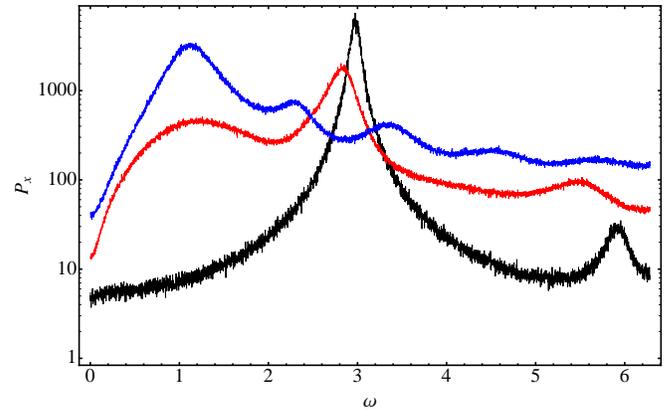}
	\caption{Power spectrum of the stochastic Brusselator averaged over $500$ realizations with $b = 9.9$ and $c = 9.0$ for system sizes $N = 10^3$(blue), $N = 10^4$(red), and $N = 10^5$(black).}
	 \label{fig-PowerSpectrumC9}
\end{figure}

We studied the intrinsically noisy Brusselator  by numerically solving the Master equation, Eq. \ref{eqn-MasterEquation}, using Gillespie's direct method \cite{Gillespie:1977ww}. Our goal was to compare the simulated dynamics with the van Kampen predictions as the system size $N$ was changed, and tuning the system through the mean-field bifurcation point.

We began by looking at the case with no time scale separation, setting $c = 1.0$ as was studied in \cite{Boland:2008}. Boland showed that the intrinsic fluctuations give rise to quasi-cycles near the mean-field bifurcation point $b_c$. These quasi-cycles are characterzed by their distinct, Lorentzian-like power spectrum. In Fig. \ref{fig-PowerSpectrumC1} we show the power spectrum at a distance $\delta = -0.1$ from the bifurcation point for three different system sizes, along with the van Kampen prediction. Each power spectrum was calculated from runs with a total time $T = 5000$ and averaged over $500$ realizations. At large $N$ the calculated power spectrum nicely matches the prediction, but as $N$ is decreased the power spectrum develops a harmonic peak at $2 \omega_0$.  Similarly, the power spectrum develops this harmonic peak as $\delta$ approaches $0$, and the quasi-cycle becomes a noisy limit cycle \cite{Boland:2008}. Note that although the van Kampen calculated power spectrum appears to decay more rapidly, all of the spectra do fall off with the same power near the peak. This can be seen more clearly by normalizing the peak heights. 
\begin{figure}[htp]
  \centering
  \begin{tabular}{cc}
    \includegraphics[width=\columnwidth]{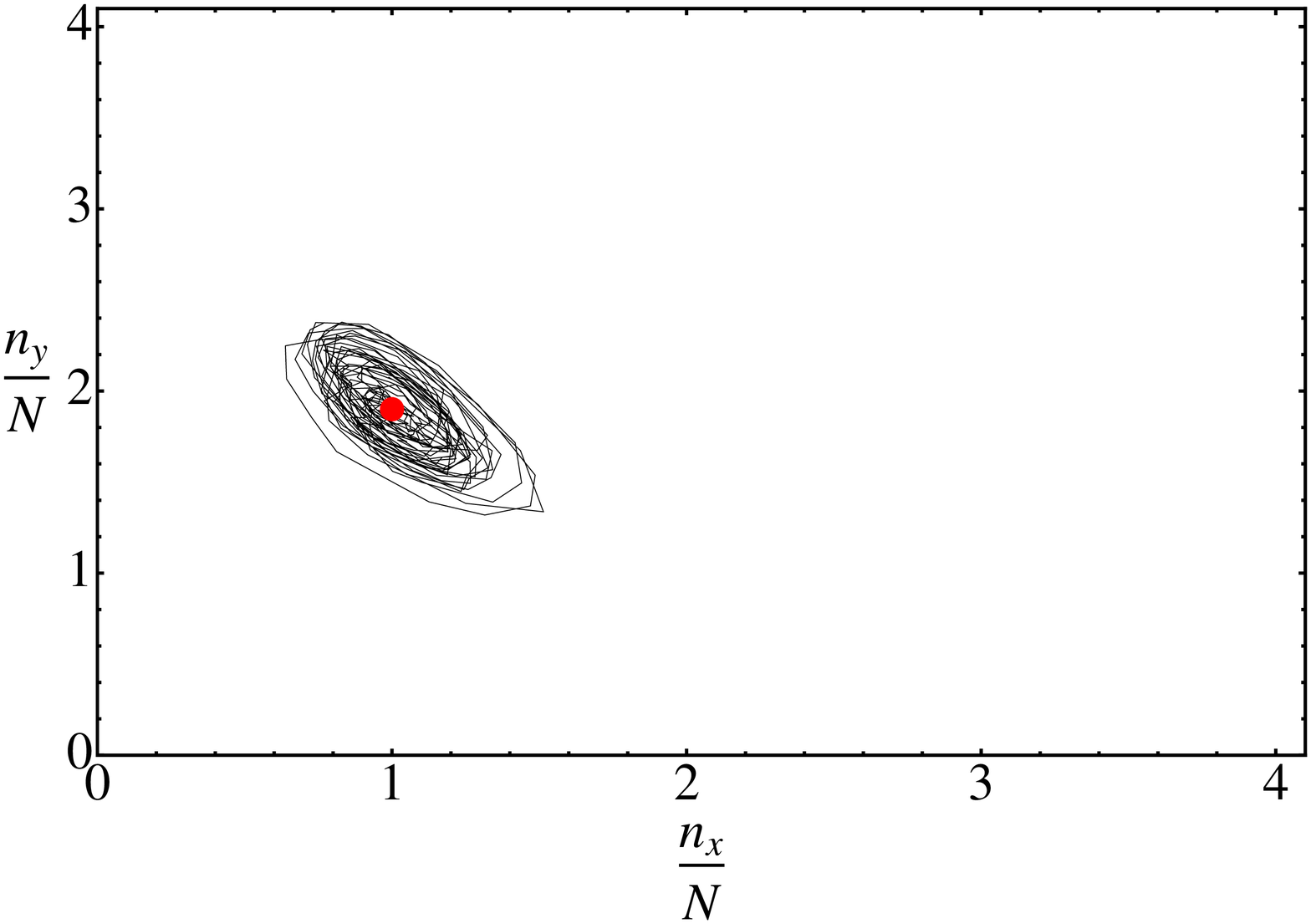}\\
    \includegraphics[width=\columnwidth]{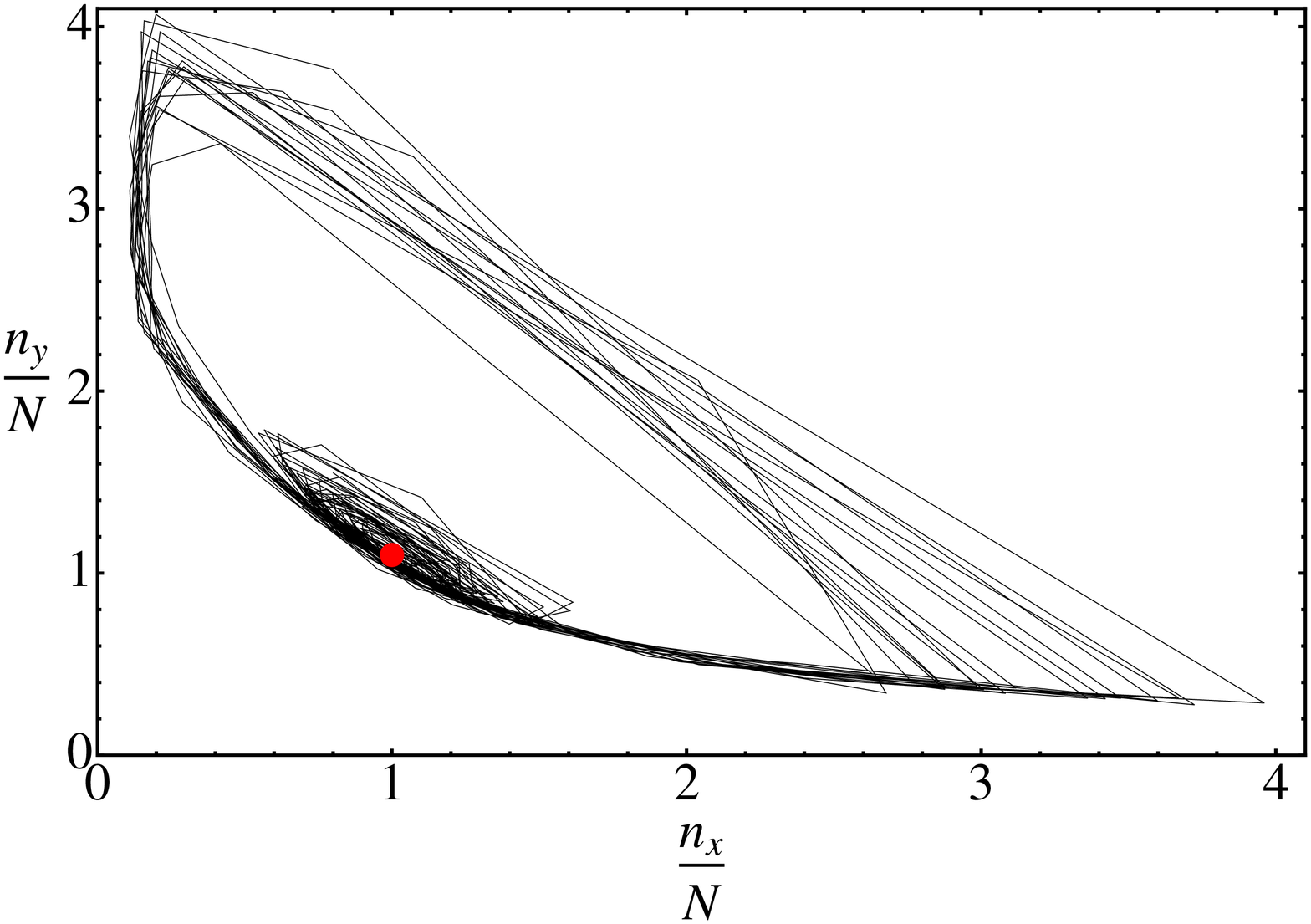}\\
  \end{tabular}
  \caption{Phase portraits for system size $N = 10^3$ at $c = 1.0, b = 1.90$ showing small amplitude quasicycles (top) and $c = 9.0, b = 9.90$ showing both small amplitude quasicycles and large amplitude excursions(bottom).}
	\label{fig-PhaseSpacePlots}
\end{figure}

As $c$ is increased, decreasing the characteristic timescale of $y$, the power spectrum undergoes a continuous change in shape dependent on the system size. From the van Kampen expansion, we expect the peak frequency to go as $c^{1/2}$, however, for smaller system sizes a peak emerges near $\omega = 1$. This can be seen in Fig. \ref{fig-PowerSpectrumC9}, where we plot the power spectra for the same $N$ and $\delta$ as were plotted in Fig. \ref{fig-PowerSpectrumC1}. Here, at $N = 10^5$, the power spectrum still matches closely to the van Kampen approximation, while at $N = 10^3$ it is dominated by this $\omega = 1$ peak.

We can begin to understand this qualitative change in behavior by looking at the simulated phase space trajectories in the two different regimes, shown in Fig. \ref{fig-PhaseSpacePlots}. Without the timescale separation, the system will  closely orbit the fixed point, shown as a red dot, where the size and shape of the orbit can be determined from the stead-state solution to the Fokker-Planck equation. The covariance of these fluctuations can be shown to depend on both $b$ and $c$, and have an overall scaling proportional to $N^{-1/2}$. When the timescale separation is introduced, the same small amplitude quasi-cycles occur around the fixed points, but there is also now a larger, almost deterministic periodic trajectory around the fixed point. In this regime, the system will wander around the fixed point until it hits the edge of the linearly stable region, at which point it will go on a large amplitude excursion before returning again to the linearly stable region. The system may also undergo multiple large amplitude cycles before returning. Comparing the stochastic phase space trajectories with the mean field ones shown in Fig. \ref{fig-MFPaseSpacePlots},  the difference in behavior can be traced back to the appearance of  a line of fixed points in the  $c \rightarrow \infty$ limit.  In that limit,  longitudinal fluctuations along the degenerate nullclines would overwhelm any transverse fluctuations, causing the system to execute the large amplitude excursions even deep in the fixed pint regime of the mean field phase diagram.  For large  but finite $c$,  transverse fluctuations result in the system entering the small amplitude quasicycle.   In this regime, therefore, we expect to see mixed mode oscillations with switching between the small amplitude quasi cycles and the large amplitude excursions.

\begin{figure}
	\includegraphics[width=\columnwidth]{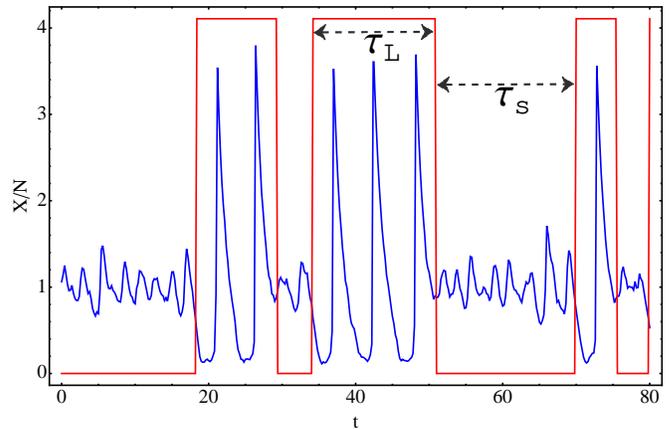}
	\caption{Typical time series displaying mixed mode oscillations for the system with parameters $N = 10^3$, $b = 9.90$, and $c = 9.0$. $\tau_L$ and $\tau_s$ denote the first passage times from large to small amplitude and small to large amplitude oscillations respectively.}
	 \label{fig-MMOTimeSeries}
\end{figure}

To investigate this behavior, we consider the transitions between oscillations of different types to be first passage processes. We define $\tau_s$ to be the first passage time going from small to large amplitude oscillations, and $\tau_L$ as the first passage time from large to small. An illustration of these times is shown on the time series plot in Fig \ref{fig-MMOTimeSeries}. We then collected first passage times at 5 different system sizes between $N = 10^3$ and $N = 10^4$ at a range of $b$ values near the bifurcation point for $c = 9.0$. The first passage time distribution for $\tau_s$ was found to be exponential, while the distribution for $\tau_L$ consisted of sharp peaks at integer multiples of the large amplitude period $T_L$ that decayed within an exponential envelope. In this case we define the mean first passage time by binning the times around each peak and using $\langle \tau_L \rangle = \sum_{n}nT_L p_n$, where $p_n$ is the weight of each peak. 
\begin{figure}[htp]
    \includegraphics[width=\columnwidth]{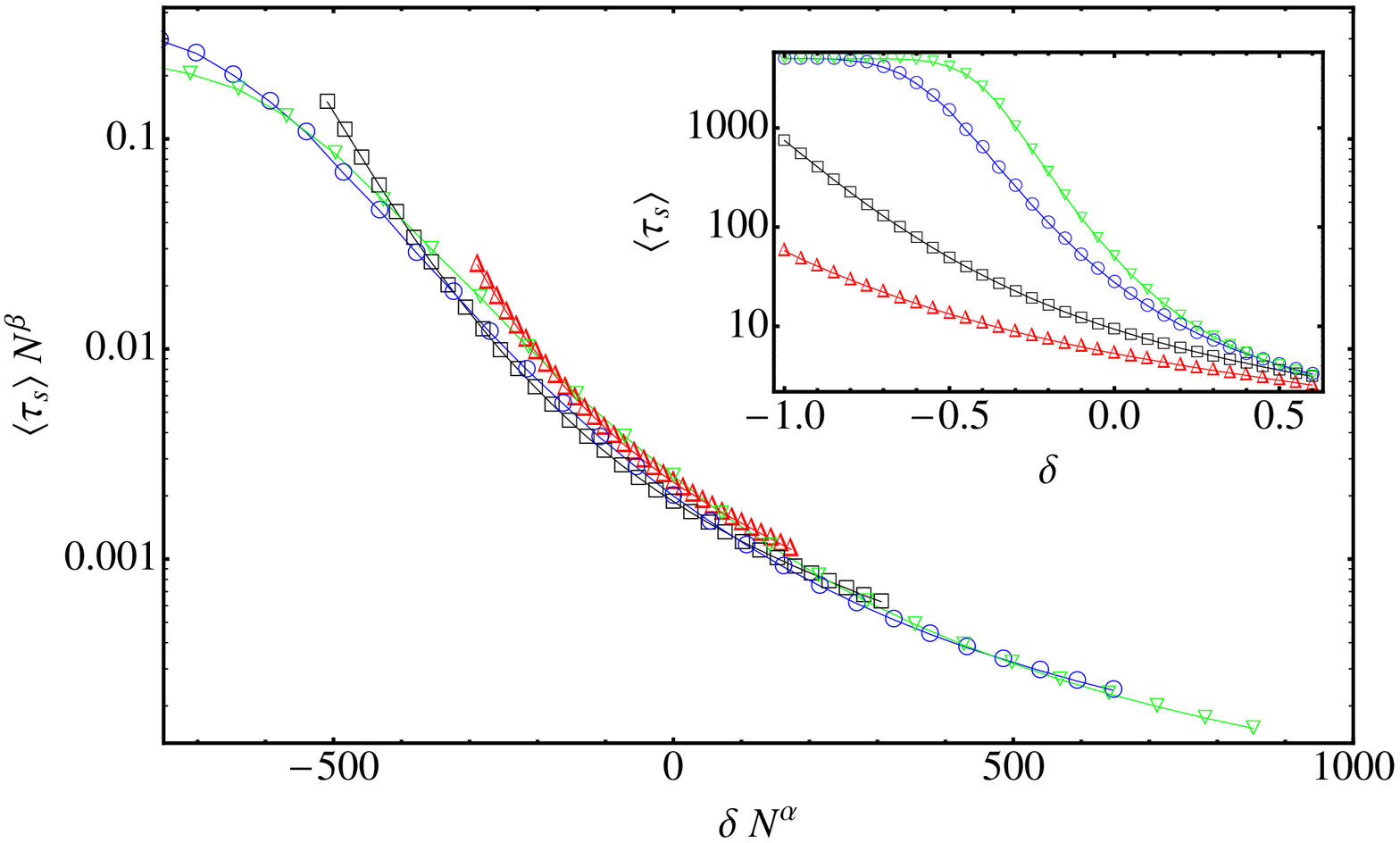}
  \caption{Average small-to-large first passage times for system sizes $N = \{10^3 (\Delta), 2\times10^3 (\square), 5\times10^3 (\circ), 7\times10^3 (\bigtriangledown)\}$ as a function of the distance from the bifurcation, $\delta$, unscaled (inset) and scaled by the system size with exponents $\alpha = 0.82$ and $\beta = 1.12$.}
	\label{fig-FirstPassageTimes-TauS}
\end{figure}
\begin{figure}[htp]
    \includegraphics[width=\columnwidth]{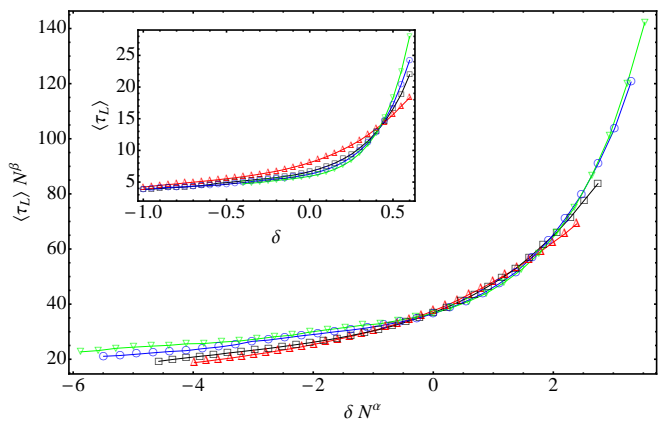}
  \caption{Average large-to-small first passage times for system sizes $N = \{10^3 (\Delta), 2\times10^3 (\square), 5\times10^3 (\circ), 7\times10^3 (\bigtriangledown)\}$ as a function of the distance from the bifurcation, $\delta$, unscaled (inset) and scaled by the system size with exponents $\alpha = \beta = 0.20$.}
	\label{fig-FirstPassageTimes-TauL}
\end{figure}

In Fig. \ref{fig-FirstPassageTimes-TauS} and Fig. \ref{fig-FirstPassageTimes-TauL} we plot the average first passage times as a function of $\delta$ at different system sizes for $\tau_s$ and $\tau_L$ respectively. As expected, $\langle \tau_s\rangle$ decreases monotonically, while $\langle \tau_L\rangle$ is monotonically increasing. The inflection points on the $\langle \tau_s\rangle$  for larger systems is due to the first passage times being of the same order as the total simulation time.

\begin{figure}
	\includegraphics[width=\columnwidth]{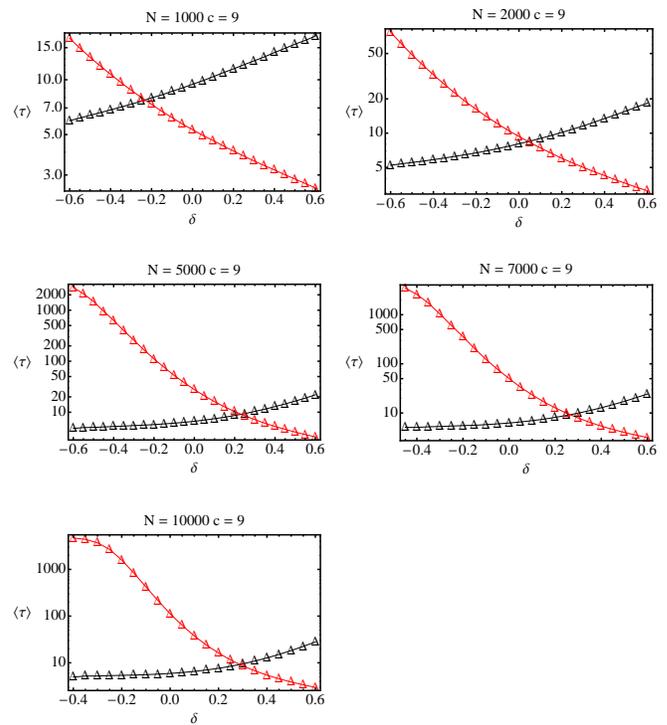}
	\caption{Average first passage times as a function of $\delta$ for 5 different systems sizes($N = 10^3, 2\times10^3, 5\times10^3, 7\times10^3, 10^4$). As the system size increases the point at which $\langle \tau_s \rangle = \langle \tau_L \rangle$ shifts to the left, approaching the point where the amplitude jumps in the mean-field singular Hopf bifurcation. }
	 \label{fig-FirstPassageCrossing}
\end{figure}

We then looked for a scaling function of the form $\langle \tau \rangle = N^{-\beta}f(\delta N^\alpha)$ to see how the first passage times depend on the system size, or noise strength. Figures \ref{fig-FirstPassageTimes-TauS} and \ref{fig-FirstPassageTimes-TauL} show the results of this scaling, where we found $\alpha _s= 0.82, \beta_s = -1.12$ for $\langle \tau_s\rangle$ and $\alpha_L = 0.2, \beta _L= 0.2$ for $\langle \tau_L\rangle$. So while both first passage times do appear to have a systematic dependence on the noise strength, it is different for each process. This is also apparent when  $\langle \tau_s\rangle$ and  $\langle \tau_L\rangle$ are plotted against one another as in Fig. \ref{fig-FirstPassageCrossing}. As the system size increases, we see that the point where the two timescales cross shifts to the point where we see an amplitude jump in the mean-field system shown in Fig. \ref{fig-Amplitude}. Beyond this value of $\delta$ the system becomes dominated by the large amplitude oscillations. It is  also interesting to note that this crossing point occurs at the same $\langle\tau\rangle$ for all of the system sizes we analyzed. The scaling relations for   $\langle \tau_L\rangle$, and $\langle \tau_s\rangle$,  imply that the scaling functions $f(x)$  decay as $x^{\beta/\alpha}$ in order for the system to have a well-defined large $N$ limit.   In this limit, $\langle \tau_L\rangle \approx \delta$, and $\langle \tau_s\rangle \approx \delta^{-1.3} $.  These predictions are consistent with the numerical results in  Fig. \ref{fig-FirstPassageTimes-TauS}.
 
In the simulations for large $c$, there is no signature of the Hopf bifurcation in the noisy system.  Instead, there is switching between small and large amplitude oscillations, with the small amplitude oscillations dominating far away from the bifurcation and the large amplitude oscillations dominating for $\delta > \delta_0$, where the mean field amplitude shows a jump due to the singular nature of the Hopf bifurcation\cite{Baer:1986}.   For finite $c$, the scaling shown in Fig. \ref{fig-FirstPassageTimes-TauS} is expected to breakdown at a large enough value of $N$, where the Van Kampen expansion becomes a good approximation.   This breakdown occurs at larger and larger $N$s as $c$ is increased and based on the nature of the mean field phase portraits, we expect that as $c \rightarrow \infty$, the mixed mode oscillations will persist for all system sizes.\\
\section{Discussion \label{sec-Discussion}}
In this work we analyzed the effects of  demographic or intrinsic noise in the zero dimensional Brusselator, a prototypical chemical oscillator representing a well-mixed system, and showed that noise can qualitatively change the nature of the system in the limit of large timescale separation between the inhibitor and the activator. We verified previous work showing that the noise induced quasi-cycles become limit cycle oscillations as the system is tuned past the bifurcations point and then showed that this transition occurs earlier for smaller systems. We then looked at what happens when a time scale separation between the two variable is introduced and found that a noise dependent switching between small and large amplitude oscillation occurs. Based on the scaling relations for the first passage times,  we speculate that carefully taking the limits $N \rightarrow\infty$ and $c \rightarrow\infty$ simultaneously will lead to a new set of deterministic equations.   This will be the subject of future work.

Previous analysis of the Brusselator in the limit where the activator time scale is much shorter than the inhibitor time scale has shown the presence of stochastic resonance\cite{Osipov:2000tr}.  Noise-induced mixed mode oscillations in systems whose mean field equations exhibit a singular Hopf bifurcation has also been demonstrated in earlier work \cite{Muratov:2008di}.  Such studies demonstrate the singular nature of noise close to the bifurcation. Our analysis of the Brusselator in the regime where the inhibitor time scale is fast indicates a singular effect of intrinsic noise in regions far away from the bifurcation.  The origin of noise-induced mixed mode oscillations in regimes where the fixed point should be stable is the emergence of a line of fixed points in the $c \rightarrow \infty$ limit.  For such a line of fixed points, noise has a large destabilizing influence similar to ``soft modes'' in classical statistical mechanics systems where fluctuations can destroy a fixed point \cite{Chaikin-Lubensky}.

We are currently working on extending the work presented here by looking at how noise-induced bistability affects the dynamics of the spatially extended Brusselator system, where diffusion is relevant. Our approach is to look at a lattice of diffusively couple point oscillator and apply the same techniques presented here. Preliminary simulations have shown similar switching behaviors, but with different first passage time distributions.
\begin{acknowledgments}
The authors acknowledge partial support of this research by Brandeis NSF-MRSEC DMR-0820492 and by the NSF-IGERT program. We would also like to acknowledge the many useful discussions with Irv Epstein, Seth Fraden, Daniel Goldstein, and Tom Butler.
\end{acknowledgments}
\bibliography{StochasticBrusselatorRefs}
\end{document}